\documentclass{ws-ijmpa}

\begin{document}

\markboth{A. Biegun for PANDA}
{CHARM PHYSICS PERFORMANCE STUDIES FOR $\overline{\textrm{P}}$ANDA}

%
\catchline{}{}{}{}{}
%

\title{CHARM PHYSICS PERFORMANCE STUDIES FOR $\overline{\textrm{P}}$ANDA
}

\author{ALEKSANDRA BIEGUN\footnote{
Present address: Delft University of Technology, Radiation Detection and Medical Imaging, Mekelweg 15, 2629 JB Delft, The Netherlands}} 

\address{KVI, University of Groningen, Groningen, The Netherlands\\
for the PANDA Collaboration
\\
a.k.biegun@tudelft.nl}

\maketitle

\begin{history}
\received{Day Month Year}
\revised{Day Month Year}
\end{history}

\begin{abstract}
The study of the charmonium ($\overline{c}c$) system is a powerful tool to understand the
strong interaction. 
In $\overline{\textrm p}$p annihilations studied with $\overline {\textrm P}$ANDA, 
the mass and width of the charmonium state, such as h$_c$, will be measured with 
an excellent accuracy, determined by the very precise knowledge of the $\overline{\textrm p}$ 
beam resolution ($\frac{\Delta p}{p}=$10$^{-4}$$-$10$^{-5}$) and not li\-mi\-ted by the resolution 
of the detector. The ana\-lysis of  h$_c$ demonstrates the feasibility to accurately determine a specific final 
state in the spectrum of charmed mesons. The preliminary background analysis of the 
$\overline{\textrm p}$p$\rightarrow\pi^{0}\pi^{0}\pi^{0}$ decay competing with a signal channel 
$\overline{\textrm p}$p$\rightarrow$h$_c$$\rightarrow\eta_{c}\gamma\rightarrow$($\pi^{0}\pi^{0}\eta$)$\gamma$Π
is under control. A comparison of three decay modes of charmonium h$_{c}$ via the electromagnetic transition is~presented.

\end{abstract}

\ccode{PACS numbers: 21.31.-x, 21.45.+v, 25.10.+s}

\section{Introduction}    

One of the main items in the broad range of the experimental program of $\overline{\textrm{P}}$ANDA 
is the charmonium spectroscopy\cite{PhysPerfRep}. 
Information about the spin-dependent interaction of heavy quarks can be obtained from a precise 
measurement of the 1$P$ hyperfine mass splitting 
$\Delta$M$_{hf}$. A non-zero hyperfine splitting may give an indication of non-vanishing spin-spin
interactions in the charmonium potential models\cite{Swanson09}.
Recently, the charmonium $\textrm h_{c}$ was studied in e$^{+}$e$^{-}$ experiments,
BESIII\cite{BES32010} and CLEO-c\cite{CLEO2008},
in the decay of $\psi$(2S)$\rightarrow\pi^{0}$h$_{c}$. 
The h$_{c}$ mass measured by BESIII, M(h$_{c}$)$=$(3525.40$\pm$0.13$\pm$0.18)~MeV/c$^{2}$, 
and branching ratios, $B$($\psi$(2S)$\rightarrow\pi^{0}$h$_{c}$)$=$(8.4$\pm$1.3$\pm$1.0)$\cdot$10$^{-4}$
and  $B$(h$_{c}\rightarrow\gamma\eta_{c}$)$=$(54.3$\pm$6.7$\pm$5.2)\%, are consistent with published CLEO 
results and they are of comparable precision. With BESIII the width of 
h$_{c}$, $\Gamma$(h$_{c}$)$=$(0.73$\pm$0.45$\pm$0.28)~MeV, was also measured.
These values can be improved by the $\overline{\textrm{P}}$ANDA experiment, where scans 
around the resonance with the high precision ($\frac{\Delta p}{p}=$10$^{-4}$$-$10$^{-5}$) anti-proton
beam will be available. Such resonance scans will allow 
to determine the total width $\Gamma$(h$_{c}$) with an accuracy of less than 0.5~MeV.

\section{$\overline{\textrm{P}}$ANDA benchmark channel: $\overline{\textrm p}$p$\rightarrow$h$_{c}\rightarrow \gamma \eta_{c}$}

The electromagnetic (EM) transition of the charmonium state h$_{c}$ to the ground state of charmonium, $\eta_{c}$,
together with different decay modes of the $\eta_{c}$,
are the most promising decay modes for the h$_{c}$ observation with 
$\overline{\textrm{P}}$ANDA\cite{PhysPerfRep,Biegun0910}. 
Examples of possible decay modes of $\eta_{c}$ with partial (for given $\eta_{c}$ decay) branching ratio (BR) 
and total BR (including BRs of sub-decays: BR$_{\phi \rightarrow K^{+}K^{-}}$$=$0.49, 
BR$_{\pi^{0} \rightarrow \gamma\gamma}$$=$0.99, 
BR$_{\eta \rightarrow \gamma\gamma}$$=$0.39) are shown in the table below.
The estimated numbers of collected events/day for the two experiment modes available 
for $\overline{\textrm{P}}$ANDA, the high luminosity mode with 
$L^{\textrm{HL}}=$2$\cdot$10$^{32}$ cm$^{-2}$s$^{-1}$
and the high resolution mode with $L^{\textrm{HR}}=$10$^{31}$ cm$^{-2}$s$^{-1}$ are presented 
together with the estimated $\textrm {S/B}$ ratios.
\begin{table}[ht]
\begin{tabular}{lcrrrrr}
                        &                    &                   &                                          &   Collected               &   Collected &  \\
Decay mode & Partial BR & Total BR  &  $\varepsilon_{Reco}$ &  events/day              & events/day& $\textrm {S/B}$\\
                        &                    &                   &                                (\%)   & ($L^{\textrm{HL}}$) &  ($L^{\textrm{HR}}$) &\\
\hline
\hline
$\eta_{c}\rightarrow \gamma\gamma$ & 4.3$\cdot$10$^{-4}$ & 4.3$\cdot$10$^{-4}$ & 8   & 20  & 1 &  $\geq$88  \\
$\eta_{c}\rightarrow$$\phi\phi$             &  2.6$\cdot$10$^{-3}$ & 6.2$\cdot$10$^{-4}$ & 24 & 92  & 4  & $\geq$ 8   \\
$\eta_{c}\rightarrow \pi^{0}\pi^{0}\eta$ &  1.6$\cdot$10$^{-2}$ & 6.3$\cdot$10$^{-3}$ & 26 & 931 & 47 & $\geq$60 \\
\hline
\end{tabular}
\end{table}

\vspace{-0.35cm}
\section{Summary}

In this work the comparison of the charmonium h$_{c}\rightarrow\gamma\eta_{c}$ decay via the EM transition is presented. 
The total BR of the $\eta_{c}\rightarrow\pi^{0}\pi^{0}\eta$ decay is the largest from all the studied decays,
which is advantageous. The preliminary analysis of the 
$\overline{\textrm p}$p$\rightarrow\pi^{0}\pi^{0}\pi^{0}$ background
with a cross section estimated to be around 4.8$\mu$b gives a $\textrm {S/B}\geq$60.
Both of these parameters show that charmonium h$_{c}$ with the decay mode of 
$\eta_{c}\rightarrow \pi^{0}\pi^{0}\eta$ is a very good candidate to be measured
with $\overline{\textrm{P}}$ANDA for a high-precision analysis.

\section*{Acknowledgments}

This research was supported by Veni-grant 680-47-120 
from the Netherlands Organization for Scientific Research (NWO), 
the University of Groningen and the Gesellschaft f\"ur Schwerionenforschung mbH (GSI), 
Darmstadt.


\begin{thebibliography}{1}    

\bibitem{PhysPerfRep} 
PANDA Collab. (W. Erni {\it et al.}),
arXiv:0903.3905v1 [hep-ex].

\bibitem{Swanson09}E. S. Swanson {\it et al.}, {\it Phys. Rep.} {\bf 429}, 243 (2006), and references therein.

\bibitem{BES32010}M. Ablikim {\it at al.}, arXiv:1002.0501v2 [hep-ex].

\bibitem{CLEO2008} S. Dobbs {\it et al.}, {\it Phys. Rev. Lett.} {\bf 101}, 182003 (2008); arXiv:0805.4599v1.

\bibitem{Biegun0910} Aleksandra Biegun for the $\overline{\textrm{P}}$ANDA collaboration,
{\it Int. J. Mod. Phys. A }{\bf 24}, 462 (2009);
A. Biegun for the $\overline{\textrm{P}}$ANDA collaboration,
{\it Acta Phys. Pol. B }{\bf 41}, 277 (2010).

\end{thebibliography}
\end{document}